\documentclass[onefignum,onetabnum]{siamonline171218}

\usepackage{lipsum}
\usepackage{amsmath,amssymb}
\usepackage{amsfonts}
\usepackage{graphicx}
\usepackage{epstopdf}
\usepackage{algorithmic}
\ifpdf
  \DeclareGraphicsExtensions{.eps,.pdf,.png,.jpg}
\else
  \DeclareGraphicsExtensions{.eps}
\fi

\usepackage{enumitem}
\setlist[enumerate]{leftmargin=.5in}
\setlist[itemize]{leftmargin=.5in}

\newsiamremark{remark}{Remark}
\newsiamremark{hypothesis}{Hypothesis}
\crefname{hypothesis}{Hypothesis}{Hypotheses}
\newsiamthm{claim}{Claim}

\headers{Bifurcations and dynamics in particle focusing}{R. Valani, B. Harding, and Y. Stokes}

\title{Bifurcations and dynamics in inertial focusing of particles in curved rectangular ducts \thanks{Submitted to the editors DATE.
\funding{This research is supported under Australian Research Council’s Discovery Projects funding scheme (project number DP160102021 and DP200100834).}}}

\author{Rahil N. Valani\thanks{School of Mathematical Sciences, University of Adelaide, South Australia, Australia 
  (\email{rahil.valani@adelaide.edu.au},\,\email{yvonne.stokes@adelaide.edu.au}).}
\and Brendan Harding\thanks{School of Mathematics and Statistics, Victoria University of Wellington, New Zealand 
  (\email{brendan.harding@vuw.ac.nz}).}
\and Yvonne M. Stokes\footnotemark[2]}

\usepackage{amsopn}

\makeatletter
\newcommand*{\addFileDependency}[1]{% argument=file name and extension
  \typeout{(#1)}% latexmk will find this if $recorder=0 (however, in that case, it will ignore #1 if it is a .aux or .pdf file etc and it exists! if it doesn't exist, it will appear in the list of dependents regardless)
  \@addtofilelist{#1}% if you want it to appear in \listfiles, not really necessary and latexmk doesn't use this
  \IfFileExists{#1}{}{\typeout{No file #1.}}% latexmk will find this message if #1 doesn't exist (yet)
}
\makeatother

\newcommand*{\myexternaldocument}[1]{%
    \externaldocument{#1}%
    \addFileDependency{#1.tex}%
    \addFileDependency{#1.aux}%
}
\ifpdf
\hypersetup{
  pdftitle={Bifurcations and dynamics of particles in inertial focusing in curved ducts with rectangular cross-section},
  pdfauthor={R. N. Valani, B. Harding, and Y. Stokes}
}
\fi

\myexternaldocument{ex_supplement}

\begin{document}

\maketitle

% REQUIRED
\begin{abstract}
%\RV{Add abstract here}
Particles suspended in fluid flow through a curved duct focus to stable equilibrium positions in the duct cross-section due to the balance of two dominant forces: (i) inertial lift force - arising from the inertia of the fluid, and (ii) secondary drag force - resulting from cross-sectional vortices induced by the curvature of the duct. Such particle focusing is exploited in various medical and industrial technologies aimed at separating particles by size. Using the theoretical model developed by Harding et al.~\cite{harding_stokes_bertozzi_2019}, we numerically investigate the dynamics of neutrally buoyant particles in fluid flow through curved ducts with rectangular cross-sections at low flow rates. We explore the rich bifurcations that take place in the particle equilibria as a function of three system parameters - particle size, duct bend radius and aspect ratio of the cross-section. We also explore the transient dynamics of particles as they focus to their equilibria by delineating the effects of these three parameters, as well as the initial location of the particle inside the cross-section, on the focusing dynamics.
\end{abstract}

% REQUIRED

% REQUIRED

\section{Introduction}

%Motion of particles suspended in a fluid flow is governed by hydrodynamic forces acting on the particles from the surrounding flow. 
Motion of particles suspended in a fluid flow is governed by the hydrodynamic interactions between the particles and the surrounding fluid. In the very low Reynolds number regime where no inertia is present, the reversibility of Stokes-flow confines the particle motion to streamlines of the background flow. For increasing Reynolds number, the inertia of the fluid can no longer be neglected and the hydrodynamic forces from the fluid can cause particles to migrate across streamlines. This phenomenon is known as inertial migration and the corresponding hydrodynamic force component that facilitates particle migration across streamlines is known as the inertial lift force. Inertial migration was first demonstrated in the classical experiment of Segr{\'e} and Silberberg~\cite{SEGRE1961} where particles suspended in flow through a straight circular pipe were observed to focus to an annular region with radius approximately $0.6$ times that of the pipe. This focusing of particles arising from inertial migration has led to a whole new class of microfluidic technologies known as inertial microfluidics. Inertial microfluidics has found many applications in medical and industrial settings such as isolation of circulating tumor cells (CTCs)~\cite{CTCs3,CTCs1,CTCs2}, separation of particles and cells~\cite{Cells2,Cells1,Cells4,Cells3}, bacteria separation~\cite{bacteria}, cell cycle synchronization~\cite{cellcycle}, flow cytometry~\cite{flowcyto}, water filtration~\cite{water1}, detection of malaria pathogen~\cite{malria}, extraction of blood plasma~\cite{plasma} and identification of small-scale pollutants in environmental samples~\cite{pollutant}. Recent developments in inertial microfluidics are highlighted in several review articles~\cite{review2,review3,Stoecklein2019,review1}. Current advances in inertial microfluidics are mainly driven by experimental trial-and-error where different designs of microfluidic devices can be assembled and tested within a few days. However, the ability to predict and optimize inertial focusing behaviors for different applications based on first principles is still work in progress~\cite{review2,Stoecklein2019}.

To predict and optimize particle focusing, it is important to understand the fundamental aspects of inertial migration physics. There have been several theoretical investigations that use analytical asymptotic methods to rationalize inertial particle migration in 2D channels~\cite{asmolov_1999,ho_leal_1974,schonberg_hinch_1989} as well as 3D channels with a circular cross-section~\cite{matas_morris_guazzelli_2004,matas_morris_guazzelli_2009}. In the case of 3D non-circular ducts, a purely analytical treatment becomes difficult and a combination of asymptotics and numerical simulations have been used. Hood et al.~\cite{phdthesishood,Hood2016,hood_lee_roper_2015} considered inertial migration of neutrally buoyant spherical particles in flow through straight ducts with rectangular cross-sections by using a combination of perturbation theory and numerical simulations. Harding et al.~\cite{harding_stokes_bertozzi_2019} extended this work and considered inertial migration of neutrally buoyant spherical particles at low flow rates in curved ducts with rectangular and trapezoidal cross-sections. It was found that wider rectangular cross-sections with aspect ratio greater than one and a trapezoidal cross-section had a better ability to separate particles based on their sizes. Moreover, the curvature of the duct plays an important role in determining the particle focusing locations. Ha et al.~\cite{Kyung2021} developed a reduced order model, the Zero Lift Fit (ZeLF) model, by fitting curves to the inertial lift force and the secondary drag force calculated from the theoretical model of Harding et al.~\cite{harding_stokes_bertozzi_2019}. The curve fitting was done such that it preserves the core topology of the two driving force fields. Using this simplified model, they investigated various dynamical behaviors of small particles and the bifurcations in the particle equilibria for flow through a curved duct with a square cross-section. Although these studies rationalize the particle focusing behavior within particular rectangular and trapezoidal channels as a function of the bend radius of the curved duct, an understanding of the bifurcations in particle equilibria and the particle focusing dynamics as a function of the other system parameters is lacking.

In this paper, we extend the work of Harding et al.~\cite{harding_stokes_bertozzi_2019} and Ha et al.~\cite{Kyung2021} by performing a systematic investigation of the bifurcations in particle equilibria for neutrally buoyant spherical particles in curved ducts having rectangular cross-sections. We also explore the transient dynamics of particles as they focus to their respective equilibria by investigating the effect of various system parameters on the focusing dynamics. The paper is organized as follows. In section~\ref{sec: theoretical model} we outline the Leading Order Force Model of Harding et al.~\cite{harding_stokes_bertozzi_2019} that has been used for our analysis. In section~\ref{sec: equilibrium focusing} we explore the bifurcations in particle equilibria as a function of three system parameters: duct bend radius, particle size and aspect ratio of the rectangular duct cross-section. In section~\ref{sec: focusing dynamics} we delineate the effects of the above three system parameters, as well as the initial location of the particle within the duct cross-section, on the focusing dynamics of the particles. We conclude in section~\ref{sec: conclusion}.

\section{Theoretical Setup}\label{sec: theoretical model}

\begin{figure}
\centering
\includegraphics[width=\columnwidth]{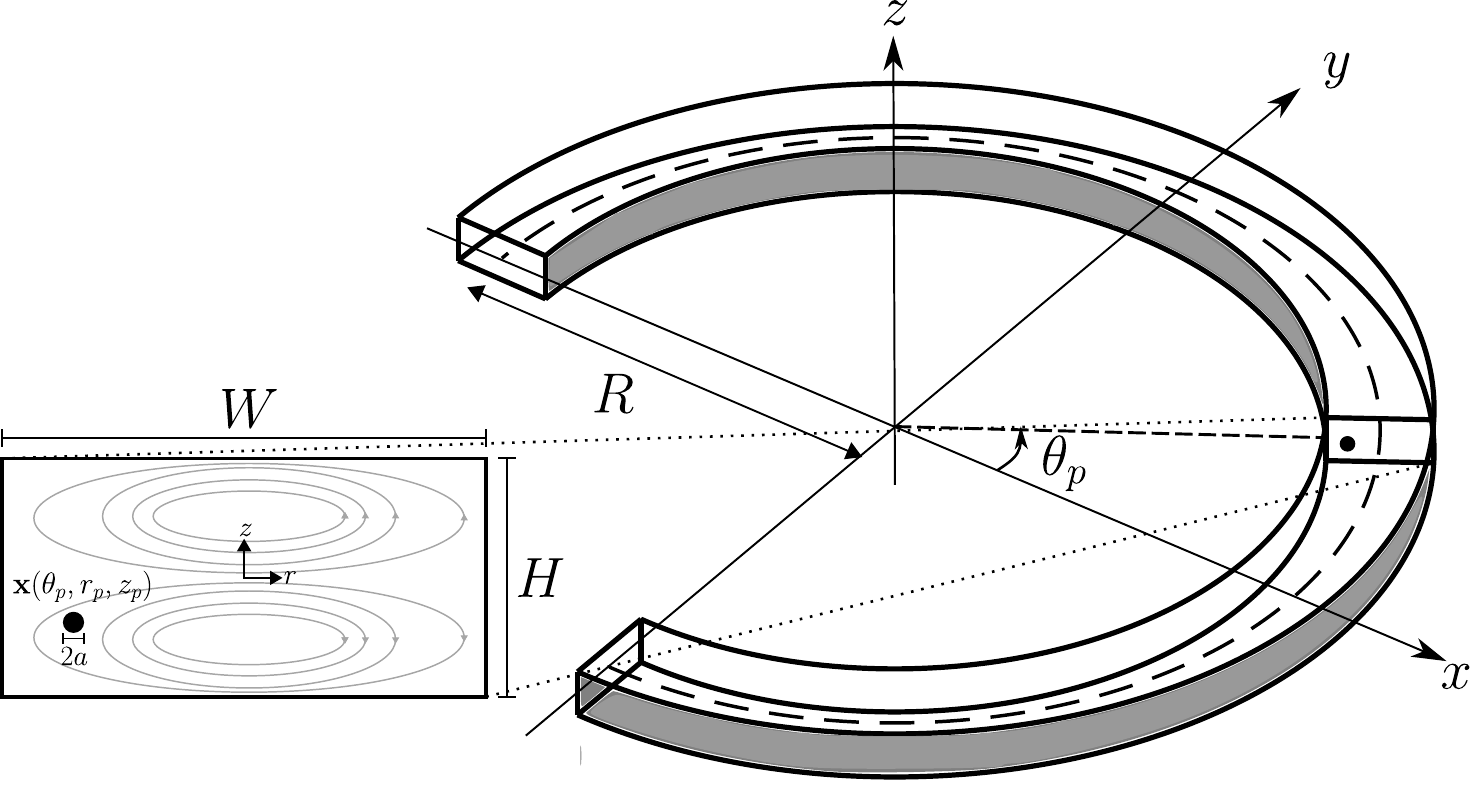}
\caption{Schematic of the theoretical setup. A particle of radius $a$ with center located at $\mathbf{x}_p=\mathbf{x}(\theta_p,r_p,z_p)$ is suspended in a fluid flow through a curved duct of radius $R$ having a rectangular cross-section of width $W$ and height $H$. The enlarged view of the cross-section illustrates the local cross-sectional $(r,z)$ co-ordinate system, and the secondary flow (gray closed curves) induced by the curvature of the duct.
%in addition to the primary axial flow along the curved duct.
}
\label{Fig: schematic}
\end{figure}

As shown in Figure~\ref{Fig: schematic}, consider a particle of radius $a$ suspended in a fluid flow through a curved duct of constant radius $R$ with a uniform cross-sectional rectangular shape of aspect ratio $AR=W/H$, where $W$ is the width and $H$ is the height of the rectangle. The horizontal and vertical co-ordinates within the rectangular cross-section are denoted by $r$ and $z$ with the origin at the center of the rectangle, and the domain is $-W/2\leq r \leq W/2$, $-H/2\leq z \leq H/2$. These cross-sectional co-ordinates are related to the global co-ordinates of the three-dimensional curved duct as follows:
\begin{equation*}
    \mathbf{x}(\theta,r,z) = (R+r)\cos(\theta) \,{\mathbf{i}}+(R+r)\sin(\theta)\,{\mathbf{j}}+z\,{\mathbf{k}}.
\end{equation*}
Here $\theta$ is the angular co-ordinate around the curved duct. The location of the particle's center is given by $\mathbf{x}_p=\mathbf{x}(\theta_p,r_p,z_p)$. We now provide an outline of the theoretical model developed by Harding et al.~\cite{harding_stokes_bertozzi_2019} which has been used for the results presented in this paper. 

\subsection{Leading Order Force model}

Harding et al.~\cite{harding_stokes_bertozzi_2019} developed a general model for the forces that govern the motion of a neutrally buoyant spherical particle in flow through a curved duct at sufficiently small flow rates. In the absence of the particle, the incompressible steady fluid flow in a curved duct driven by a steady pressure gradient is referred to as Dean flow~\cite{Dean1927,Dean1959} and has been studied extensively for ducts with a rectangular cross-section~\cite{winters_1987,Yamamoto_2004}. The pressure gradient drives an axial flow in the curved duct which results in the development of a secondary flow in the cross-section consisting of counter-rotating vortex pairs (typically one pair) due to the centrifugal force on fluid parcels as they flow around the bend. The presence of a particle disturbs this background Dean flow and the resulting dimensionless equations for the cross-sectional disturbance pressure $q$ and the disturbance fluid velocity $\mathbf{v}$ in a constant rate rotating frame (moving with the particle) are given by
\begin{align}\label{Eq: NS}
- \nabla q + \nabla^2\mathbf{v}&=\text{Re}_p \left( \left(\mathbf{v}+\bar{\mathbf{u}}-\Theta \left(\mathbf{e}_{z}\times \mathbf{x}\right)\right)\cdot \nabla \mathbf{v}+\mathbf{v}\cdot\nabla \bar{\mathbf{u}} +\Theta\left(\mathbf{e}_z\times\mathbf{v}\right)\right) \:\:\: \text{on}\:\:\mathbf{x} \in \mathcal{F},\\ \nonumber
\nabla \cdot \mathbf{v} &= 0 \:\:\: \text{on}\:\:\mathbf{x} \in \mathcal{F},\\ \nonumber
\mathbf{v}&=\mathbf{0} \:\:\: \text{on}\:\:\mathbf{x} \in \partial \mathcal{D},\\ \nonumber
\mathbf{v}&=-\bar{\mathbf{u}}+\Theta\left(\mathbf{e}_z\times\mathbf{x}\right)+\mathbf{\Omega}_p\times\left(\mathbf{x}-\mathbf{x}_p\right) \:\:\: \text{on}\:\:\mathbf{x} \in \partial \mathcal{F} \setminus \partial \mathcal{D}.
\end{align}
Here, $\bar{\mathbf{u}}$ is the background fluid flow velocity in the absence of the particle, $\mathbf{\Omega}_p$ is the particle spin and $\mathbf{e}_{z}$ is the unit vector in the vertical $z$ direction, all in a reference frame which is rotating about the $z$ axis at a constant rate $\Theta:=\partial \theta_p/\partial t$. Moreover, $\mathcal{D}$ denotes the interior of the duct, $\partial \mathcal{D}$ denotes the boundaries of the duct, $\mathcal{F}:=\{\mathbf{x}\in\mathcal{D}:|\mathbf{x}-\mathbf{x}_p|\geq 1\}$ is the dimensionless fluid domain in the presence of the particle and $\partial \mathcal{F} \setminus \partial \mathcal{D}=\{\mathbf{x}:|\mathbf{x}-\mathbf{x}_p|=1\}$ is the surface of the particle. The particle Reynolds number is, $\text{Re}_p=\rho U_m a^2/(\mu l)$, {where} $\rho$ {is} the density of the particle/fluid, $\mu$ is the fluid dynamic viscosity and $U_m$ is a characteristic velocity of the background fluid flow. The variables have been non-dimensionalized using the following scales: the particle radius $a$ for $\mathbf{x}$, $l/U_m$ for time $t$, $U_m a/l$ for the velocities $\mathbf{v}$ and $\bar{\mathbf{u}}$, $U_m/l$ for the particle spin $\mathbf{\Omega}_p$ and the angular velocity $\Theta$, and $\mu U_m/l$ for the disturbance pressure $q$. The dimensionless force (scaled by $\rho U_m^2 a^4/l^2$) and torque (scaled by $\rho U_m^2 a^5/l^2$) acting on the particle in the rotating frame are given by
\begin{align}\label{eq: force particle}
\mathbf{F}=&-\frac{4\pi}{3}\Theta^2 \left(\mathbf{e}_z\times\left( \mathbf{e}_z\times \mathbf{x}_p \right) \right)+ \int_{|\mathbf{x}-\mathbf{x}_p|<1} \bar{\mathbf{u}}\cdot\nabla\bar{\mathbf{u}}\,\text{d}V \\ \nonumber
&+\frac{1}{\text{Re}_p} \int_{|\mathbf{x}-\mathbf{x}_p|=1} (-\mathbf{n})\cdot(-q\mathbf{I}+\nabla\mathbf{v}+\nabla\mathbf{v}^T)\,\text{d}S,
\end{align}
\begin{align}\label{eq: torque particle}
\:\:\:\:\:\:\:\:\:\:\:\:\:\:\:\:\:\:\:\:\:\mathbf{T}=&-\frac{8\pi}{15}\Theta \left(\mathbf{e}_z\times\mathbf{\Omega}_p \right)+ \int_{|\mathbf{x}-\mathbf{x}_p|<1} (\mathbf{x}-\mathbf{x}_p)\times(\bar{\mathbf{u}}\cdot\nabla\bar{\mathbf{u}})\,\text{d}V \\ \nonumber
&+\frac{1}{\text{Re}_p} \int_{|\mathbf{x}-\mathbf{x}_p|=1} (\mathbf{x}-\mathbf{x}_p)\times((-\mathbf{n})\cdot(-q\mathbf{I}+\nabla\mathbf{v}+\nabla\mathbf{v}^T))\,\text{d}S.  
\end{align}
Performing a perturbation expansion of the disturbance flow in powers of the particle Reynolds number,
\begin{align*}
\mathbf{v} &= \mathbf{v}_0+\text{Re}_p \mathbf{v}_1+{O}(\text{Re}^2_p), \\ \nonumber
q &= q_0+\text{Re}_p q_1+{O}(\text{Re}^2_p),    
\end{align*}
and substituting into \eqref{Eq: NS} gives us a zeroth order system {for} $q_0,\mathbf{v}_0$ and a first order system {for} $q_1,\mathbf{v}_1$. The zeroth order system captures all  the non-zero boundary conditions and the first order system captures the most significant inertial contribution to the complete equations.
Substituting the perturbation expansions in \eqref{eq: force particle} and \eqref{eq: torque particle} gives
\begin{align*}
\mathbf{F}&=\text{Re}_p^{-1}\mathbf{F}_{-1}+\mathbf{F}_{0}+{O}(\text{Re}_p), \\ \nonumber
\mathbf{T}&=\text{Re}_p^{-1}\mathbf{T}_{-1}+\mathbf{T}_{0}+{O}(\text{Re}_p).
\end{align*}
In order to understand separately how the axial and secondary flow components affect the particle motion, the variables are separated into the axial component (denoted by subscript `$a$') and the secondary component (denoted by subscript `$s$'). This, after some algebra and approximations (see Harding et al.~\cite{harding_stokes_bertozzi_2019}) results in the following equation for the cross-sectional force on the particle: 
%\ys{(can the subscript $p$ on $\bf F$ below be removed; it is confusing in the context of the just introduced subscripts and is not before used for the force)}:
\begin{equation*}
    \mathbf{F}_s=(\mathbf{e}_r\cdot(\text{Re}^{-1}_p\mathbf{F}_{-1,s}+\mathbf{F}_0)\mathbf{e}_r+(\mathbf{e}_z\cdot(\text{Re}^{-1}_p\mathbf{F}_{-1,s}+\mathbf{F}_0))\mathbf{e}_z.
\end{equation*}
Using this force, we can construct a first order model for the trajectory of the particle. This gives,
\begin{equation*}
    \frac{\text{d} r_p}{\text{d} t}=-\text{Re}_p\frac{F_{p,r}}{C_r},\:\:\:
    \frac{\text{d} z_p}{\text{d} t}=-\text{Re}_p\frac{F_{p,z}}{C_z}\:\:\:\text{and}\:\:\:\frac{\text{d} \theta_p}{\text{d} t}=\frac{\bar{{u}}_a}{R/a+r_p},
\end{equation*}
where $\bar{\mathbf{u}}_a$ is the axial component of the background fluid flow, $F_{s,r}=\mathbf{F}_s\cdot\mathbf{e}_r$ and $F_{s,z}=\mathbf{F}_s\cdot\mathbf{e}_z$ are the radial and the vertical components of the cross-sectional force, respectively, with corresponding drag coefficients $C_r$ and $C_z$ that vary with the particle's position in the cross-section. 
%\ys{(More subscripts need definition as the secondary component is now separated into radial and vertical components; perhaps the $p$ subscript can be removed.)}

Numerical implementation of this model involves using a Finite Element Method to compute the forces acting on the particle. We refer the reader to Harding et al.~\cite{harding_stokes_bertozzi_2019} for additional details on the numerical method. Once the forces are pre-computed at numerous points in the cross-section and for numerous system parameter values, an interpolant of $C_r$, $C_z$, $F_{s,r}$, $F_{s,z}$ is constructed and the particle dynamics are then simulated using the MATLAB solvers ode45 and ode15s.

To find the equilibrium positions of the particle, i.e. the fixed points of the particle flow field within the cross-section, we find locations $(r^*,z^*)$ where $F_{s,r}=F_{s,z}=0$. The stability of the equilibrium points is determined by calculating the eigenvalues $\lambda$ of the following Jacobian matrix:
\begin{equation*}
    J=\begin{bmatrix}
\partial F_{s,r}/\partial r & \partial F_{s,r}/\partial z\\
\partial F_{s,z}/\partial r & \partial F_{s,z}/\partial z
\end{bmatrix}.
\end{equation*}
Here the derivatives are evaluated at the fixed point $(r^*,z^*)$. The nature of the eigenvalues determine the type of particle equilibrium.

For the results presented in this paper, we use the following non-dimensional scales for the system variables: dimensionless bend radius $\tilde{R}=2R/H$, dimensionless particle size $\tilde{a}=2a/H$, dimensionless time $\tilde{t}=U_m t/H$ and cross-sectional co-ordinates $\tilde{r}=2r/H$ and $\tilde{z}=2z/H$.

\section{Bifurcations in particle focusing equilibria}\label{sec: equilibrium focusing}

In this section, we explore the bifurcations taking place in the equilibrium positions of the particle (i.e. where the net cross-sectional force on the particle is zero, $F_{s,r}=F_{s,z}=0$) as a function of the system parameters. We do this by considering three different cross-sections: square {($AR=1$)}, $2\times1$ rectangular {($AR=2$)} and $1\times2$ rectangular {($AR=1/2$)}, and investigate the bifurcations in particle equilibria as a function of the dimensionless particle size $\tilde{a}$ and the dimensionless bend radius $\tilde{R}$. We also explore the variations in the stable equilibria of the particle inside high aspect ratio ducts for different particle sizes over a range of bend radii that are typically encountered in inertial microfluidics experiments.

\subsection{Bifurcations in a square cross-section}\label{sec: Bif square}

\begin{figure}
\centering
\includegraphics[width=0.98\columnwidth]{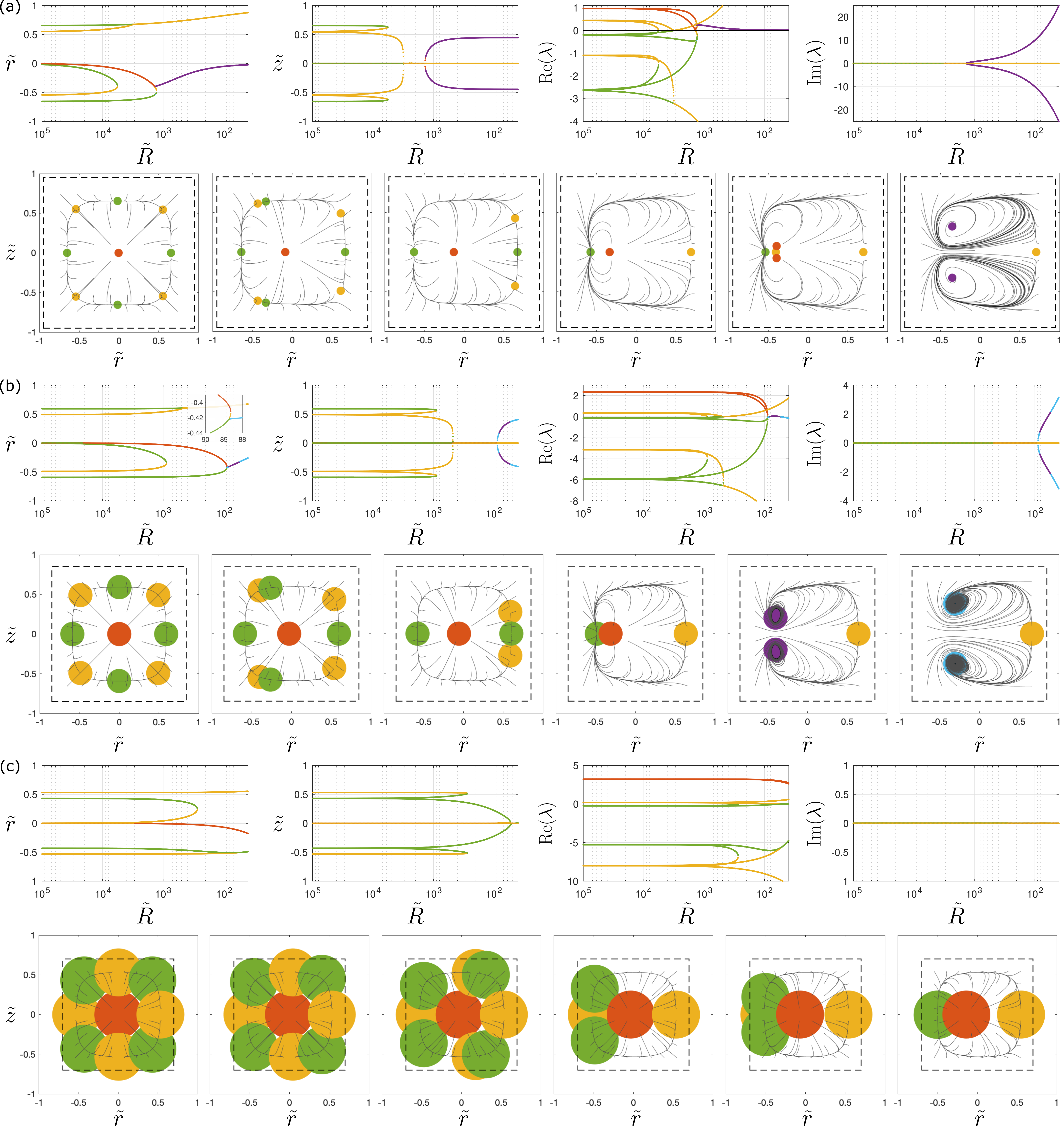}
\caption{Bifurcations in particle equilibria in a square cross-section as a function of the bend radius $\tilde{R}$ for particle sizes (a) $\tilde{a}=0.05$, (b) $\tilde{a}=0.15$ and (c) $\tilde{a}=0.30$. The radial $\tilde{r}$ and vertical $\tilde{z}$ location of the equilibria as well as the real $\text{Re}(\lambda)$ and imaginary $\text{Im}(\lambda)$ parts of the eigenvalues of the corresponding equilibria are plotted as a function of the bend radius $\tilde{R}${; note that $\tilde R$ decreases from left to right}. The panels below these bifurcation plots show the location of equilibria in the square cross-section as filled circles. The size of the circle corresponds to particle size and the color of the circles denotes the type of equilibria: unstable node in red, stable node in green, saddle point in yellow, unstable spiral in purple and a stable spiral in cyan. For panel (a) the cross-sectional images correspond to (left to right) $\tilde{R}=10^5, 6000, 4000, 1600, 1370, 1000$, for panel (b) they correspond to $\tilde{R}=10^5, 1000, 500, 100, 80, 50$, and for panel (c) they correspond to $\tilde{R}=10^5, 1000, 300, 200, 100, 50$. The gray curves in each of these images indicate the typical trajectories of particles within the cross-section while the dashed square %represent the 
{shows} locations of the center of the particle for which %the particle
{it} will hit the walls of the duct.}
\label{Fig: sq_bif}
\end{figure}

We start by exploring the bifurcations taking place in the equilibrium positions of a particle in a curved duct with a square cross-section. Figure~\ref{Fig: sq_bif} shows the cross-sectional locations as well as the real and imaginary parts of the eigenvalues for the particle equilibra as a function of the duct bend radius $\tilde{R}$ for three representative particle sizes: $\tilde{a}=0.05$, $0.15$ and $0.30$.

For all three particle sizes, we observe several equilibria in the cross-section that change in both location and classification as the duct bend radius is varied. For relatively large bend radii, where inertial lift forces dominate secondary drag forces, we observe stable and unstable nodes. For relatively small bend radii, where secondary drag forces dominate inertial lift forces, we see the emergence of stable or unstable spirals. Saddle points are observed for both large and small bend radii. 

For a small particle, $\tilde{a}=0.05$, and large bend radii, we find nine particle equilibria inside the square cross-section. Four stable nodes (green) near the center of the four edges of the square, four saddle points (yellow) near the corners of the square and an unstable node (red) near the center of the square (see Figure~\ref{Fig: sq_bif}(a)). A slow manifold is formed along a closed curve that connects all the stable nodes and saddle points. The slow manifold emerges due to a large disparity in magnitude of the two eigenvalues for each of the stable nodes and the saddle points. For each of these equilibria, the eigenvalue in a direction tangential to the slow manifold is much smaller than the eigenvalue in a normal direction, resulting in slow migration along this manifold. As the bend radius is decreased progressively, a number of bifurcations take place. Firstly, saddle-node bifurcations take place at relatively large bend radii where the stable nodes near the center of the top and bottom edges of the square collide and annihilate with the saddle points located to their left. Immediately after the saddle-node bifurcations, a critical slowing down of the particle motion is observed in the vicinity of this bifurcation location along the slow manifold. Further decrease in the bend radius results in the stable node near the center of the right side undergoing a subcritical pitchfork bifurcation with the two saddle points located above and below. The three equilibria merge into a single saddle point. As the bend radius is further reduced, the unstable node near the center of the duct migrates towards the stable node near the center of the left edge. As the two points get closer, the unstable node undergoes a supercritical pitchfork bifurcation and turns into a saddle point and produces two additional unstable nodes; one on each side in the vertical direction. These two unstable nodes change into unstable spirals (purple) with further decrease in bend radius, while the newly formed saddle point merges with the stable node in another saddle-node bifurcation. The unstable spirals develop an encompassing limit cycle around themselves. {As the bend radius is reduced to very small values,} %For very small bend radii, 
the unstable spirals with their limit cycles %migrated 
migrate towards the center of the upper and lower halves of the cross-section, while the saddle point slowly migrates towards the right edge of the square. We note that similar bifurcations were observed by Ha et al.~\cite{Kyung2021} for a square cross-section using their simplified ZeLF model.

For an intermediate sized particle of $\tilde{a}=0.15$ in a square cross-section, we find similar bifurcations to that of a small particle except at small bend radii. As shown in Figure~\ref{Fig: sq_bif}(b), at relatively small bend radii, it appears that an unstable and a stable node near the left edge merge directly and produce two unstable spirals{,} %;
one on each side in the vertical direction. However, performing numerical simulations with finer resolution near this bifurcation, it appears that the stable node near the center of the left edge first undergoes a supercritical pitchfork bifurcation, producing a saddle point and two stable nodes on either side of the saddle point in the vertical direction. The newly formed saddle point then merges with the unstable node in a saddle-node bifurcation, while the stable nodes immediately transition to stable spirals. This is shown in the inset of Figure~\ref{Fig: sq_bif}(b) where the saddle-node bifurcation is clearly visible, while the accurate resolution of the transition from stable nodes to stable spirals, which happens over a very narrow region of bend radii, is beyond the scope of the present work. The stable spirals further undergo a subcritical Hopf bifurcation and turn into unstable spirals with an encompassing limit cycle for a narrow range of bend radii before turning back into stable spirals via a supercritical Hopf bifurcation at small bend radii.

For a large particle of size $\tilde{a}=0.30$ in a square cross-section, we find a qualitative change in the particle equilibria as shown in Figure~\ref{Fig: sq_bif}(c). At large bend radii, we now find that the stable nodes and the saddle-points switch their stability. The equilibria near the corners are now stable nodes while the equilibria near the center of edges are now saddle points. As the bend radius is progressively decreased, the saddle points near the center of top and bottom edges now undergo saddle-node bifurcations with the stable nodes located to their right. Further reduction in the bend radius results in stable nodes near the top-left and bottom-left corners undergoing a subcritical pitchfork bifurcation with the saddle point at the center of the left edge, leaving behind a stable node near the center of the left edge. As the bend radius is further decreased beyond the practical range typically encountered in experimental setups ($\tilde{R} \leq 40$, not shown in Figure~\ref{Fig: sq_bif}(c)), this stable node undergoes a supercritical pitchfork bifurcation and turns into a saddle point along with two stable nodes, each located either side vertically. The saddle point  merges with the unstable node in a saddle-node bifurcation while the stable nodes turn into stable spirals at very small bend radii.

\begin{figure}[!t]
\centering
\includegraphics[width=\columnwidth]{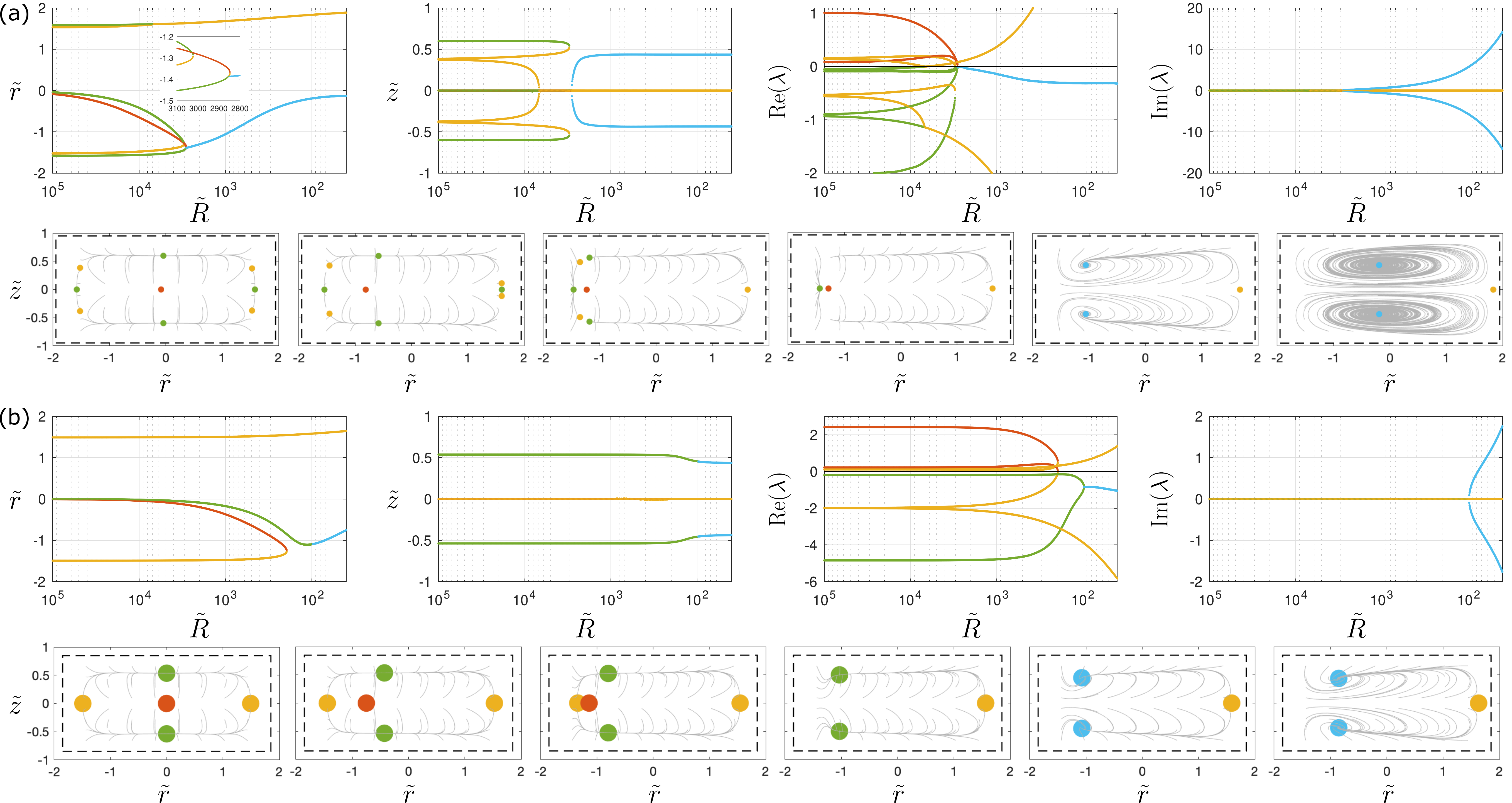}
\caption{Bifurcations in particle equilibria in a $2\times1$ rectangular cross-section as a function of the bend radius $\tilde{R}$ for particle sizes (a) $\tilde{a}=0.05$ and (b) $\tilde{a}=0.15$. The radial $\tilde{r}$ and vertical $\tilde{z}$ location of the various equilibria as well as the real and imaginary parts of the eigenvalues of the corresponding equilibria are shown as a function of $\tilde{R}$; note that $\tilde R$ decreases from left to right. The panels below these bifurcation plots show the location of equilibria in the $2\times1$ rectangular cross-section as filled circles. The size of the circle corresponds to particle size and the color of the circles denotes the type of equilibria:  unstable node in red, stable node in green, saddle point in yellow and a stable spiral in cyan. For panel (a) the cross-sectional images correspond to (left to right) $\tilde{R}=10^5, 7000, 3200, 3000, 1000, 100$, while for panel (b) they correspond to $\tilde{R}=10^5, 400, 210, 150, 90, 50$. The gray curves in each of these images indicate the typical trajectories of particles within the cross-section while the dashed square %represent the 
{shows} locations of the center of the particle for which %the particle
{it} will hit the walls of the duct.}
\label{Fig: rect_2x1}
\end{figure}

\subsection{Bifurcations in a $2\times1$ rectangular cross-section}\label{sec: Bif rect 2x1}

Figure~\ref{Fig: rect_2x1} shows the bifurcations in particle equilibria in a $2\times1$ rectangular cross-section as a function of the bend radius $\tilde{R}$ for two representative particle sizes, $\tilde{a}=0.05$ and $\tilde{a}=0.15$. %(see Supplemental material for plots of additional particle sizes).

For the smaller particle, as shown in Figure~\ref{Fig: rect_2x1}(a), we first observe a subcritical pitchfork bifurcation where the stable node near the center of the right edge merges with the two saddle points near the top-right and bottom-right corners, leaving behind a single saddle point. As the bend radius is decreased, the nodes at the center of the top and bottom edges undergo saddle-node bifurcations with the saddle points to their left. We note that although these two bifurcations are similar to what was observed for the same size particle in a square cross-section, the order of bifurcations are reversed. Further decreasing the bend radius, it appears that the unstable node at the center drifts left, and collides with the stable node and gives out two stable spirals. However, resolving this bifurcation numerically, it appears that the stable node first undergoes a subcritical pitchfork bifurcation producing two stable nodes in the vertical direction with a saddle point in between. The saddle point then merges with the unstable node while the two stable nodes change into stable spirals. This is depicted in the inset of Figure~\ref{Fig: rect_2x1}(a) where the saddle-node bifurcation is clearly visible, while the accurate resolution of the transition from stable nodes to stable spirals, which happens over a very narrow region of bend radii, is beyond the scope of the present work. The two stable spirals migrate towards the center of the duct at small bend radii while the saddle point drifts to the right edge of the duct.

\begin{figure}
\centering
\includegraphics[width=0.94\columnwidth]{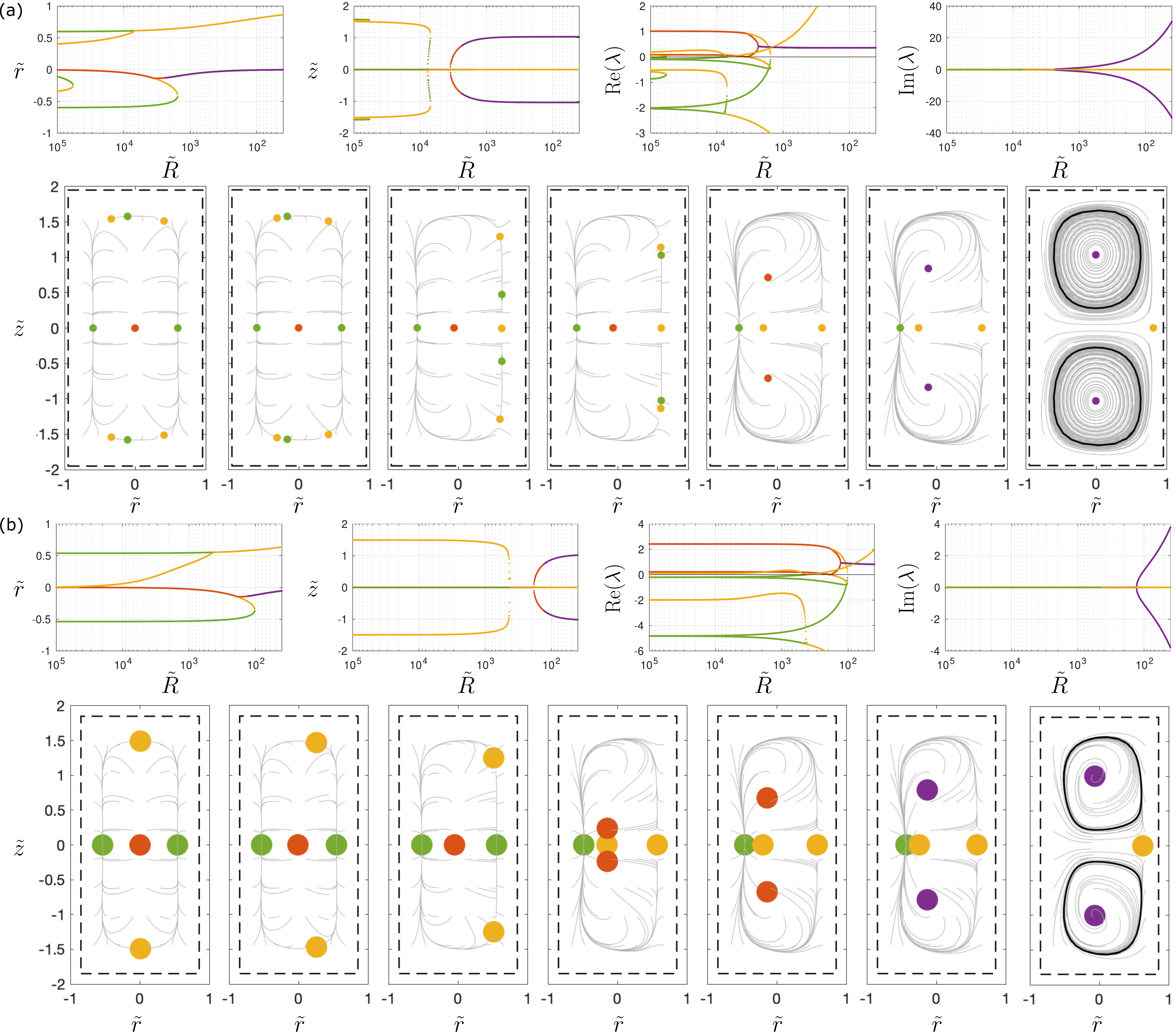}
\caption{Bifurcations in particle equilibria in a $1\times2$ rectangular cross-section as a function of the bend radius $\tilde{R}$ for particle sizes (a) $\tilde{a}=0.05$ and (b) $\tilde{a}=0.15$. The radial $\tilde{r}$ and vertical $\tilde{z}$ location of the various equilibria as well as the real and imaginary parts of the eigenvalues of the corresponding equilibria are shown as a function of $\tilde{R}$; note that $\tilde R$ decreases from left to right. The panels below these bifurcation plots show the location of equilibria in the $1\times2$ rectangular cross-section as filled circles. The size of the circle correspond to particle size and the color of the circles denote the type of equilibria:  unstable node in red, stable node in green, saddle point in yellow and an unstable spiral in purple. For panel (a) the cross-sectional images correspond to (left to right) $\tilde{R}=10^5, 70000, 7500, 7000, 2500, 2000, 100$, while for panel (b), they correspond to $\tilde{R}=10^5, 2000, 500, 180, 140, 120, 50$. The gray curves in each of these images indicate the typical trajectories of particles within the cross-section while the dashed square %represent the 
shows locations of the center of the particle for which it will hit the walls of the duct. The black curve for unstable spirals represents the limit cycle.}
\label{Fig: rect_1x2}
\end{figure}

For the larger particle, we find qualitatively different kinds of bifurcations compared to the smaller particle. Initially, at very large bend radii, we find that there are only five fixed points: an unstable node at the center, two stable nodes each near the center of top and bottom edges, and two saddle points each near the center of the left and right edges. As the bend radius is progressively decreased, the unstable node migrates to the left and undergoes a saddle-node bifurcation with the saddle point at the center of the left side. The two stable nodes first migrate to the left and change into stable spirals. The two stable spirals then migrate back towards the center of the upper and lower halves of the duct at small bend radii, while the saddle point on the right side all along drifts closer to the right edge of the duct.

\subsection{Bifurcations in a $1\times2$ rectangular cross-section}\label{sec: Bif rect 1x2}

Figure~\ref{Fig: rect_1x2} shows the bifurcations in particle equilibria in a $1\times2$ rectangular duct as a function of the bend radius $\tilde{R}$ for two representative particle sizes, $\tilde{a}=0.05$ and $\tilde{a}=0.15$.

For the smaller particle, similar to a square-cross section, we first observe saddle-node bifurcations between the two stable nodes at the top and bottom and the saddle points on their left, at relatively large bend radii. As the bend radius is decreased, unlike the square cross-section, the stable node at the center of the right edge undergoes a supercritical pitchfork bifurcation and turns into a saddle point and releases two additional stable nodes. These newly formed stable nodes undergo saddle-node bifurcations with the two saddle points on the right side. As the bend radius is further decreased, the unstable node at the center migrates left and undergoes a supercritical pitchfork bifurcation where it turns into a saddle point and releases two unstable nodes vertically. The saddle point then goes on to merge with the stable node at the center of the left edge in a saddle-node bifurcation while the two new unstable nodes turn into unstable spirals and develop encompassing limit cycles. Thus, at small bend radii, we have three fixed points: a pair of unstable spirals with limit cycles surrounding them and a saddle point on the right side.

For a larger particle, similar to a $2\times1$ rectangular cross-section, we observe only five fixed points at very large bend radii: an unstable node at the center, two stable nodes, each near the center of left and right edges, and two saddle points, each near the center of top and bottom edges. With decreasing bend radius, the two saddle points near the top and bottom edges migrate to the right and undergo a subcritical pitchfork bifurcation with the stable node on the right. This results in the three fixed points merging into a single saddle point. The unstable node migrates to the left and undergoes a supercritical pitchfork bifurcation where it turns into a saddle point and releases two unstable nodes. The saddle then goes on to merge with the stable node on the left edge, while the two unstable nodes turn into unstable spirals and develop encompassing limit cycles.

\subsection{Parameter space exploration for particle separation}\label{sec: PS space}

\begin{figure}
\centering
\includegraphics[width=\columnwidth]{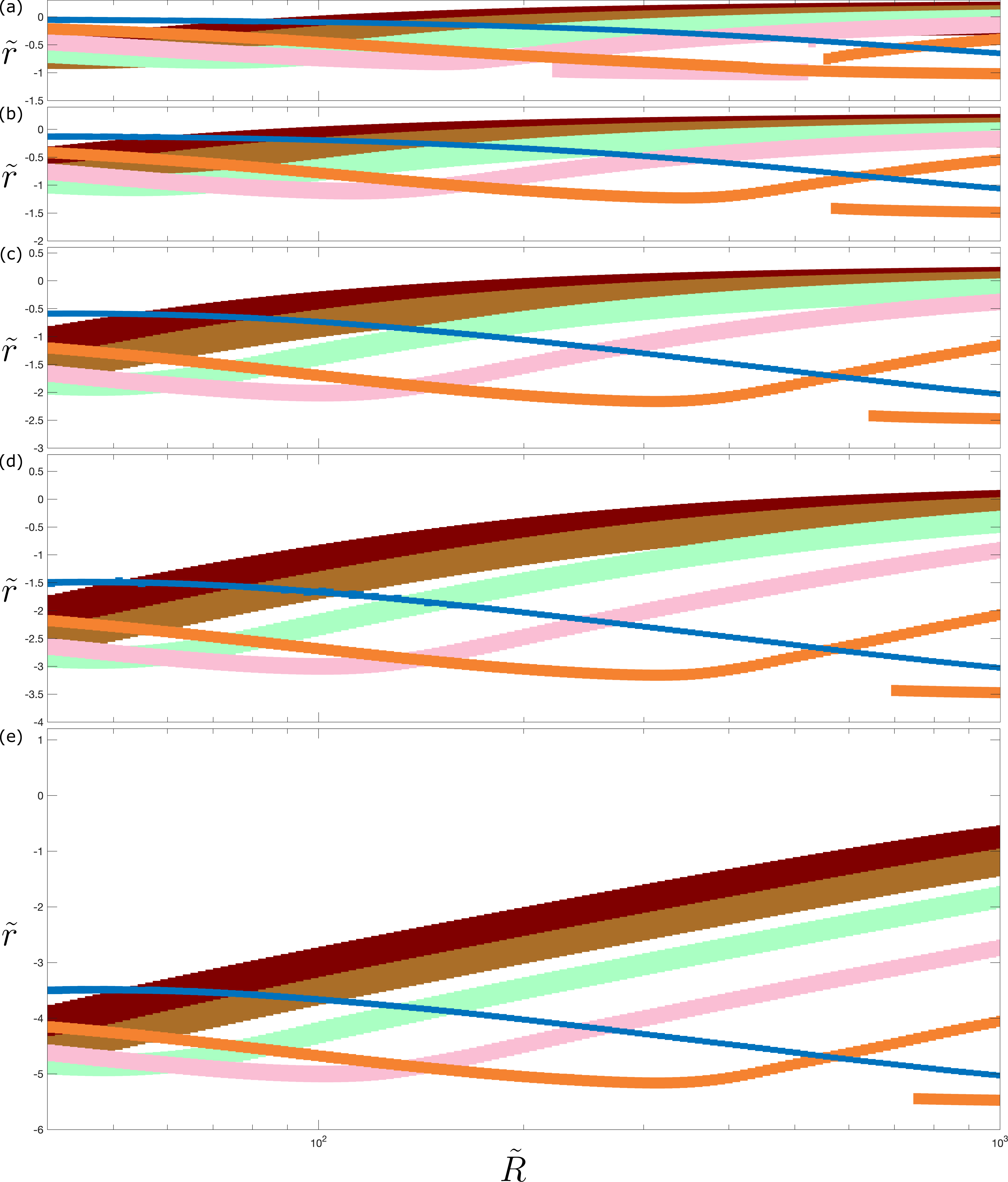}
\caption{ Stable particle focusing position in the radial $\tilde{r}$ direction as a function of the bend radius $\tilde{R}$ for different aspect ratio rectangular ducts: (a) $3\times2$, (b) $2\times1$, (c) $3\times1$, (d) $4\times1$ and (e) $6\times1$. The different colored curves represent different particles sizes: $\tilde{a}=0.05$ in blue, $\tilde{a}=0.10$ in orange, $\tilde{a}=0.15$ in pink, $\tilde{a}=0.20$ in mint green, $\tilde{a}=0.25$ in brown and $\tilde{a}=0.30$ in maroon. The width of the curves correspond to the diameter $2\tilde{a}$ of the particle. Note that the range of the vertical axis differs in each case and does not capture the full duct width.}
\label{Fig: PS space}
\end{figure}

%\begin{enumerate}
%    \item Here I can include parameter space plots where I can show the evolution of the fixed points in the parameter space of the Aspect Ratio (AR) of the cross-section and the bend radius $R$ or $\kappa$. Brendan is running simulations for AR $3 \times 2$ and $3 \times 1$ and maybe we can also do some simulations for $1 \times 2$, $1 \times 4$, $ 1\times3 $ and $2\times 3$. This will allow us to understand some general trends of the bifurcations in rectangular cross sections. This will also provide us with a big picture of where particle focusing in more pronounced in this parameter space of AR and $R$ or $\kappa$.
%\end{enumerate}

To understand the effect of the aspect ratio of the rectangular cross-sectional geometry on particle focusing, we plot the positions of stable equilibria for different aspect ratios and particle sizes as a function of the dimensionless bend radius $\tilde{R}$, as shown in Figure~\ref{Fig: PS space}. Here the radial location $\tilde{r}$ of the stable focusing positions for different sized particles is shown in the parameter space spanned by the aspect ratio and bend radius over a practical range of bend radii that are typically realized in experiments.

It is clear from Figure~\ref{Fig: PS space} that, assuming negligible particle-particle interactions, wider rectangular ducts with a larger aspect ratio are better at separation of various particle sizes compared to the smaller aspect ratio rectangular ducts. Moreover, for most of the range of bend radii shown ($\tilde{R}\lesssim 500$), the different size particles only have a single stable focusing point in the $\tilde{r}$ co-ordinate, except for the $3\times2$ cross-section. For all aspect ratios, we particularly find a good separation between the three smallest particle sizes of $\tilde{a}=0.05$, $0.10$ and $0.15$ over a broad range of bend radii. These three smallest particles also separate well from any one of the three largest particle sizes. The third largest particle, $\tilde{a}=0.20$, separates well at a large aspect ratio of $AR=6$, while we don't see a clear separation between the two largest particles for any aspect ratio. 
%As we increase the aspect ratio, the spacing between the focused particles increases and hence we get better separation. However, at high aspect ratios, the dynamics of the particles become very slow on the slow manifold and it takes longer for the particle to focus to its equilibrium position. Hence, an intermediate aspect ratio may be optimal for particle focusing in finite time where the particles are sufficiently separated in a reasonable amount of time (oq equivalently reasonable number of turns).
Hence, although we observe that it is difficult to get a good separation between all different particle sizes, we can get a good separation between two or three different particle sizes using a duct with aspect ratio of $AR=2$ or higher. In typical experiments of inertial microfluidics aimed at isolating tumor cells, the larger CTCs ($15-20\,\mu$m in diameter) are needed to be separated from the relatively smaller WBCs ($10-15\,\mu$m in diameter) and RBCs ($3-8\,\mu$m in diameter)~\cite{Warkiani2016}. Hence, for example, if the channel geometry is chosen such that $H=100\,\mu$m then we have, $\tilde{a}=0.03-0.08$ for RBCs, $\tilde{a}=0.10-0.15$ WBCs  and $\tilde{a}=0.15-0.20$ for CTCs. Hence one may be able to separate the three particles based on the locations of their stable equilibria, and the analysis presented in this paper might be useful in designing inertial microfluidic devices to separate such particles. 

%We also find a similarity in the focusing curves of different size particles at larger aspect ratios as shown in Figure.~\ref{Fig: PS space}(d-e). Here it may be possible to obtain a scaling relationship that allows us to predict the equilibrium focusing positions at larger aspect ratios based on the known positions of the particles at smaller aspect ratio. \RV{Try to do some simple scaling here to see if we can get anything analytically that allows us the predict the focusing position of particles as well as maybe the focusing time scale as the aspect ratio is increased.}
%
%\RV{Are we able to describe the transition between the various equilibrium positions more intuitively for different aspect ratio as a balance between the lift and drag forces?}
%\RV{Highlight the point that the locations in the plot are equilibrium focusing locations. Mention how the current designs of the experimental microfluidic ducts are limited in the sense that it may not giving the particls enough time to focus onto the stable equilibrium positions. Maybe propose a design .e.g a helical channel where the bend radius is constant and the particle can be run long enough to focus. Also end this section with a sentence to investigate the dynamics of particle focusing while will be explored in the next section. This will give a smooth transition.}

\section{Dynamics of focusing to equilibrium points}\label{sec: focusing dynamics}

\begin{figure}
\centering
\includegraphics[width=\columnwidth]{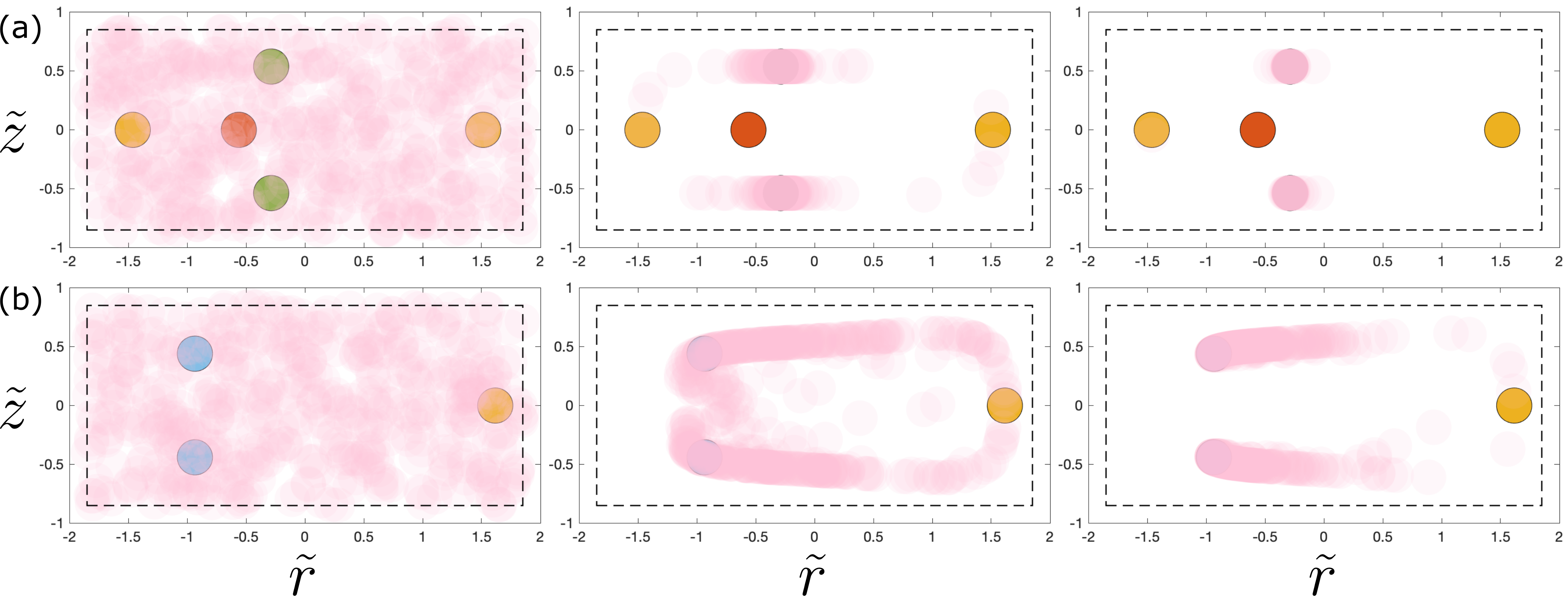}
\caption{Focusing dynamics as a function of the bend radii. Dynamics of $500$ non-interacting particles (transparent pink circles) of size $\tilde{a}=0.15$ that are initiated randomly in the $2\times1$ rectangular cross-section. Snapshots of the particles along with their equilibria are shown for bend radii (a) $\tilde{R}=600$ and (b) $\tilde{R}=60$ for three angular positions of the duct: $\theta=0^{\circ}$ (left panels), $\theta=180^{\circ}$ (middle panels) and $\theta=360^{\circ}$ (right panels). For the larger bend radius duct, these angular locations correspond to an average transit time of $\tilde{t}=0$, $395$ and $776$ (averaged over $500$ particles), while for the smaller bend radius, these correspond to an average transit time of $\tilde{t}=0$, $40$ and $78$. The filled circles denote the type of equilibria: unstable node in red, stable node in green, saddle point in yellow and a stable spiral in cyan.}
\label{Fig: dynamics focusing}
\end{figure}

While designing experimental microfluidic devices based on the principles of inertial particle focusing, in addition to knowing the location of particle equilibria and their bifurcations, it is also important to understand the dynamics of focusing to these equilibria. For example, in microfluidic devices with circular or spiral duct geometries, the dynamics will provide valuable information about the time scale of focusing to the equilibria which can then be used to calculate the required number of turns of the curved duct for particles to focus to their stable equilibria. Moreover, small particles at relatively small bend radii in low aspect ratio rectangular ducts, focus onto limit cycles and hence inertial focusing to a fixed location in the cross-section is not realized for such particles. 
%Moreover, many experimental setups exploit the dynamics of particle focusing to separate different size particles before they focus to their stable equilibrium positions~\cite{}. 
Therefore, an understanding of the particle dynamics is crucial to optimally designing the inertial microfluidic devices for particle separation. In this section, we highlight some aspects of the particle dynamics and investigate how the focusing dynamics vary as a function of the system parameters.

\begin{figure}
\centering
\includegraphics[width=\columnwidth]{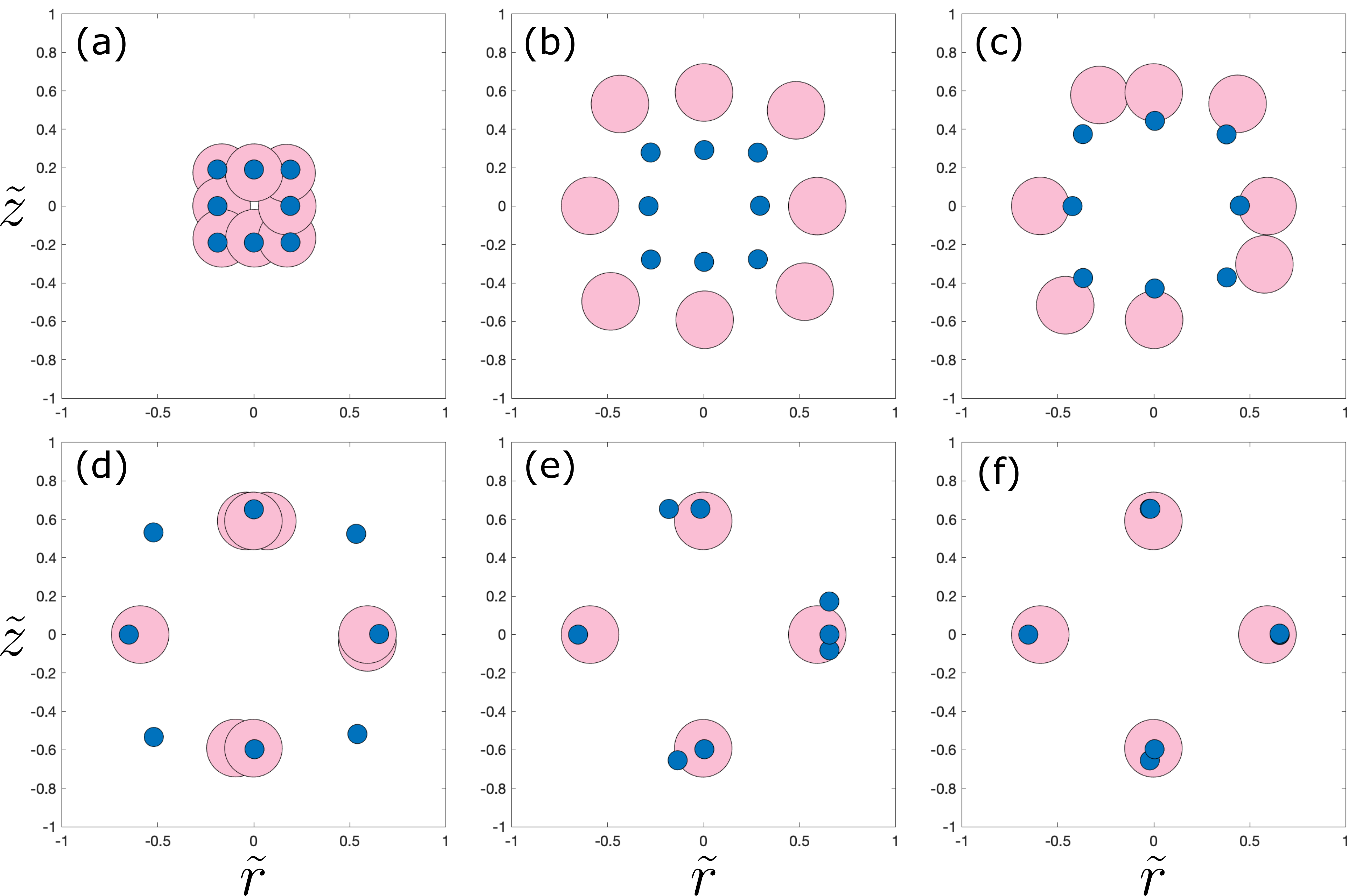}
\caption{Focusing dynamics as a function of particle size in a square cross-section at a large bend radius of $\tilde{R}=10^5$. Particle dynamics for two different sized particles, $\tilde{a}=0.05$ (blue) and $\tilde{a}=0.15$ (pink), that were initiated at similar location in the cross-section of the duct. Snapshots of the particles are shown at angular locations (a) $\theta=0^{\circ}$, (b) $0.36^{\circ}$, (c) $0.72^{\circ}$, (d) $1.8^{\circ}$, (e) $9^{\circ}$, and (f) $18^{\circ}$.}
\label{Fig: dynamics focusing 2}
\end{figure}

\subsection{Focusing dynamics with bend radius variations}

We start by exploring the effect of bend radii on the focusing dynamics. Figure~\ref{Fig: dynamics focusing} shows the evolution of $500$ initially randomly distributed non-interacting particles of the same size at two different representative bend radii, one large and one small. For the larger bend radii, we observe that the particles quickly `snap' onto a slow manifold and then slowly migrate towards the stable nodes. For the smaller bend radii, we observe that the particles quickly focus to the stable spirals. Comparing the evolution of the particle dynamics for one turn of the duct reveals that at large bend radius, most of the particles focus to their equilibria by the end of the turn, however, for the smaller bend radius, particles have not focused to the equilibria by the end of a full turn. Since the characteristic axial particle velocity is similar at large and small bend radius and the particles in the larger bend radius duct need to travel further to complete one full turn, it takes longer for the particles in the larger bend radius duct to complete one full turn compared to the particles in the smaller bend radius duct. Thus, particles in the larger bend radius duct are able to focus completely in this longer time it takes to cover one full turn. Hence, to get complete focusing of particles at stable equilibria, one typically needs multiple turns for smaller bend radii ducts. However, we note that for a smaller bend radius duct, the effect of secondary flow further enhances focusing, so the total distance (and time) can be less than that of a larger bend radius duct. Hence, while designing circular, helical or spiral microfluidic devices, one may need to optimize between the required number of turns (better at large bend radii) and the time taken to focus (better at small bend radii).

\subsection{Focusing dynamics with particle size variations}

Figure~\ref{Fig: dynamics focusing 2} shows the focusing dynamics of two different sized particles, $\tilde{a}=0.05$ and $\tilde{a}=0.15$, in a square cross-section at a relatively large bend radius. By starting the two different sized particles at similar initial locations, we observe that the bigger particle migrates to the slow manifold at a shorter angular distance compared to the smaller particle. Once both the particles are constrained onto the slow manifold, we again observe that the bigger particle is able to focus at the final stable equilibrium position using a shorter angular distance compared to the smaller particle. Hence, although we do not see a large variation in the focusing dynamics of these two different size particles which have similar equilibria and started with similar initial conditions, the focusing of the larger particle requires a smaller angular distance of the duct compared to a smaller particle.

\begin{figure}
\centering
\includegraphics[width=\columnwidth]{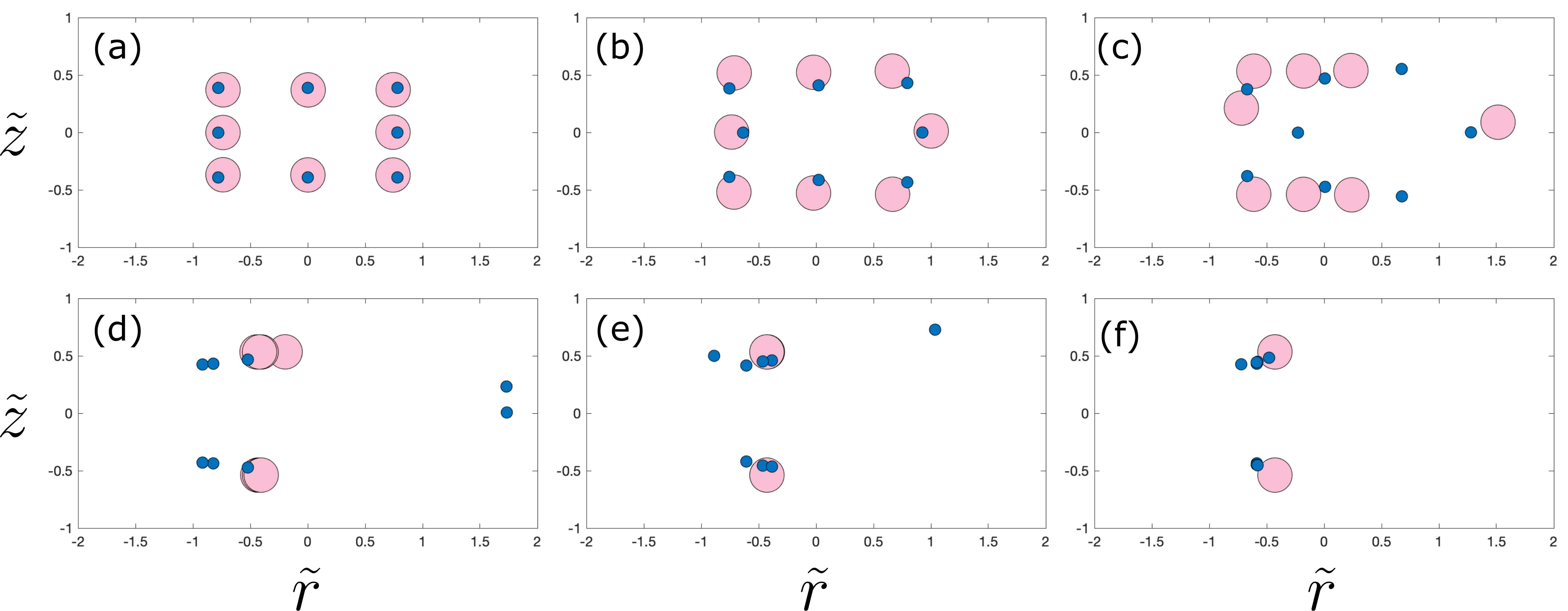}
\caption{Focusing dynamics as a function of particle size in a $2\times1$ rectangular cross-section at a moderate bend radius of $\tilde{R}=400$. Particle dynamics for two different sized particles, $\tilde{a}=0.05$ (blue) and $\tilde{a}=0.15$ (pink), that were initiated at similar location in the cross-section of the duct. Snapshots of the particles are shown at angular locations (a) $\theta=0^{\circ}$, (b) $14.4^{\circ}$, (c) $57.6^{\circ}$, (d) $360^{\circ}$, (e) $720^{\circ}$, and (f) $1440^{\circ}$.}
\label{Fig: dynamics focusing 2 2}
\end{figure}

Comparing the dynamics of the same two particle sizes in a $2\times1$ rectangular cross-section at a moderate bend radius we also find a similar behavior (see Figure~\ref{Fig: dynamics focusing 2 2}). These two different sized particles, that initially start at similar locations, focus to stable spirals. The larger particle reaches the stable spiral by traveling a shorter angular distance compared to the smaller particle.

\subsection{Focusing dynamics with aspect ratio variations}

\begin{figure}
\centering
\includegraphics[width=\columnwidth]{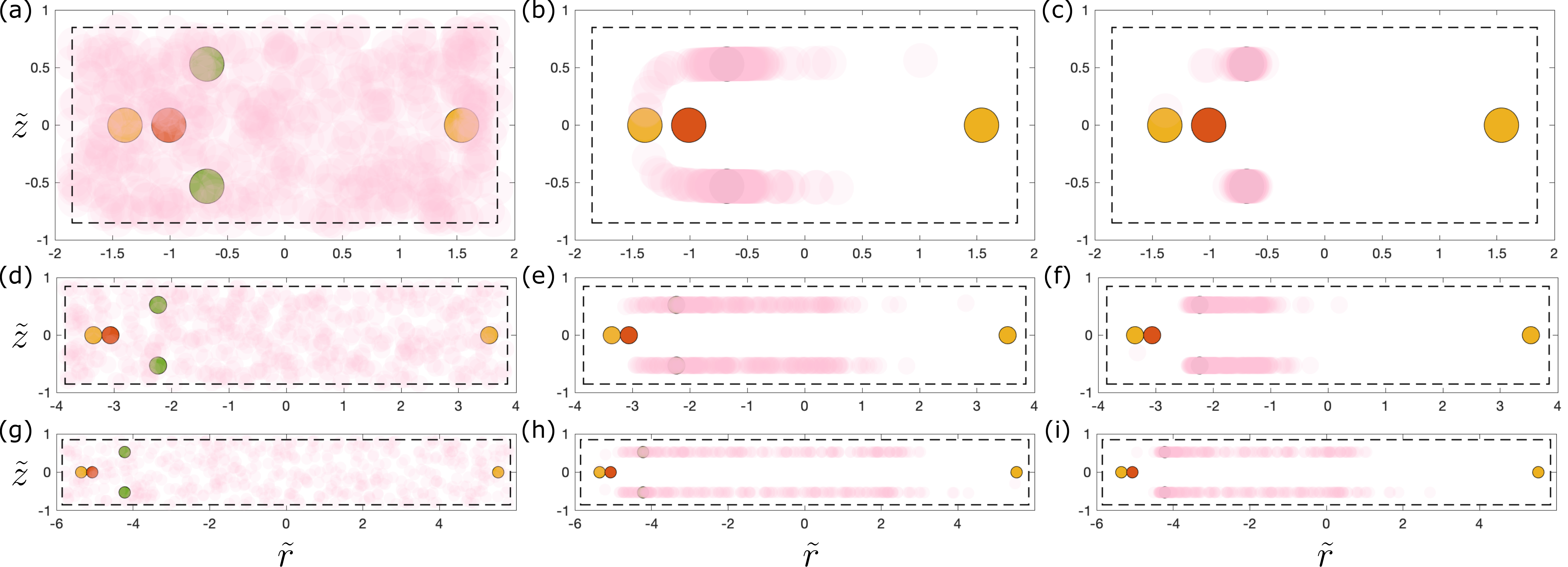}
\caption{Focusing dynamics as a function of aspect ratio. Snapshots of particle locations for randomly initiated $500$ non-interacting particles of size $\tilde{a}=0.15$ (transparent pink circles) along with their equilibria are shown at a dimensionless bend radius of $\tilde{R}=250$ for aspect ratios (a)-(c) $2\times1$, (d)-(f) $4\times1$ and (g)-(i) $6\times1$. Snapshots are shown at angular positions: (a),(d),(g) $\theta=0^{\circ}$, (b),(e),(h) $360^{\circ}$ and (c),(f),(i) $720^{\circ}$. The filled circles denote the type of equilibria: unstable node in red, stable node in green and a saddle point in yellow.}
\label{Fig: dynamics focusing 3}
\end{figure}

Figure~\ref{Fig: dynamics focusing 3} shows the focusing dynamics of $500$ initially randomly located non-interacting particles of size $\tilde{a}=0.15$ in a rectangular cross-sections with aspect ratios $2\times1$, $4\times1$ and $6\times1$. We observe that the lower aspect ratio $2\times1$ cross-section is able to focus most of the particles using a smaller angular distance compared to the two larger aspect ratio cross-sections. As shown in Figure~\ref{Fig: dynamics focusing 3}, most of the particles focus to their stable equilibria in two turns in a $2\times1$ cross-section while the $4\times1$ and $6\times1$ cross-sections are unable to give complete focusing of particles in the same two turns. Two effects contribute to this: (i) for larger aspect ratios, the particles have a longer cross-sectional distance to travel along the slow manifold to reach their equilibria, and (ii) the horizontal component of the inertial lift force diminishes near the center at larger aspect ratios which further slows down the dynamics of particles along the slow manifold. Thus, while designing inertial microfluidic devices, one may need to optimize between the quality of particle separation (better at high aspect ratios) and the number of turns required or the time taken to fully focus (better at low aspect ratios).

\subsection{Focusing dynamics with variations in initial particle location inside the cross-section}

\begin{figure}
\centering
\includegraphics[width=\columnwidth]{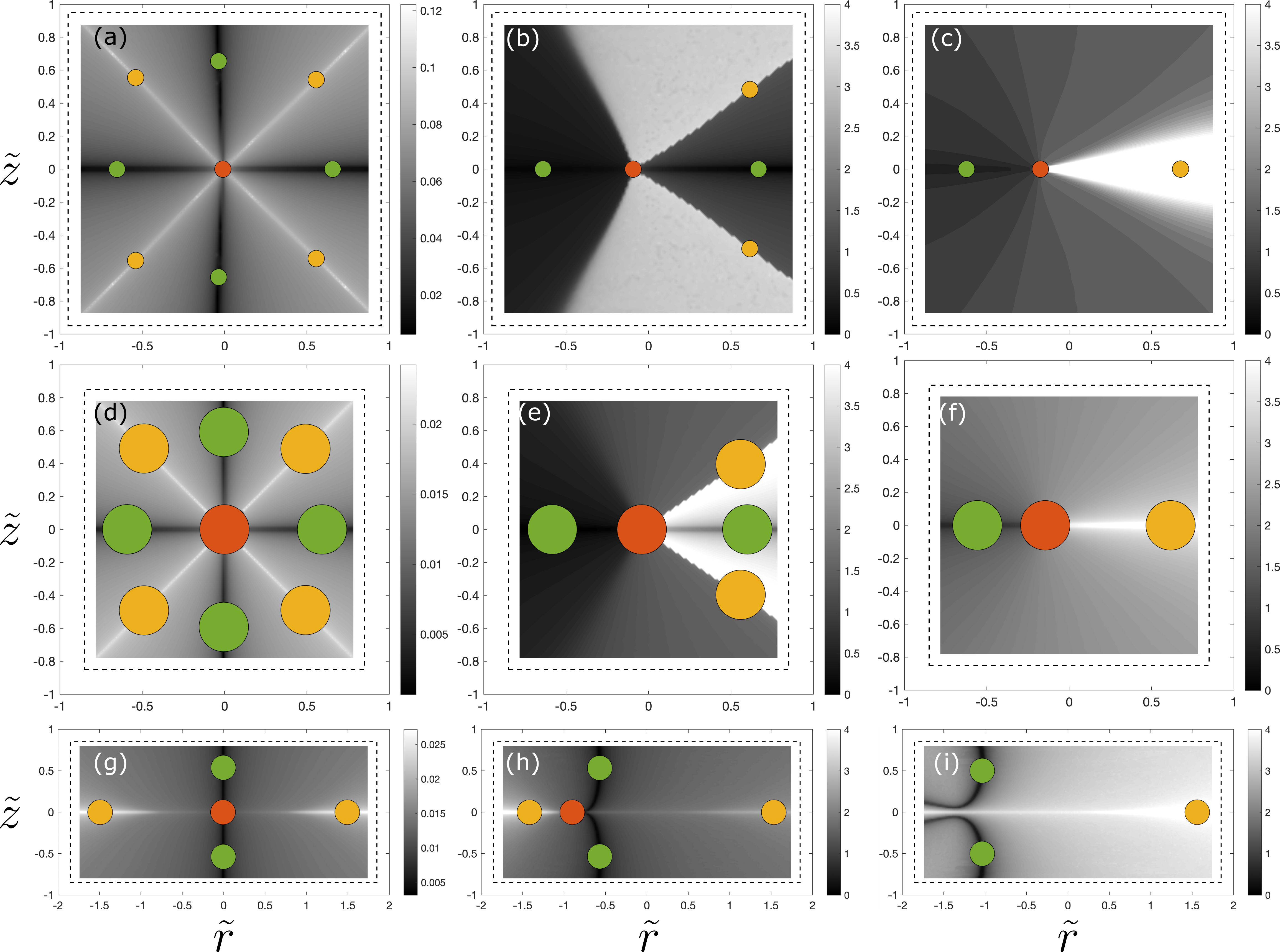}
\caption{Focusing dynamics as a function of the initial particle position within the cross-section. Filled contour map showing the number of turns, $N_{\theta}=\theta_p/2\pi$, required for the particle to focus to an equilibrium position based on its initial location within the cross-section. Particles of size $\tilde{a}=0.05$ in a square cross-section at bend radii (a) $\tilde{R}=50000$, (b) $\tilde{R}=5600$ and (c) $\tilde{R}=3000$. Particles of size $\tilde{a}=0.15$ in a square cross-section at bend radii (d) $\tilde{R}=50000$, (e) $\tilde{R}=700$ and (f) $\tilde{R}=200$. Particles of size $\tilde{a}=0.15$ in a rectangular $2\times1$ cross-section at bend radii (g) $\tilde{R}=50000$, (h) $\tilde{R}=300$ and (i) $\tilde{R}=150$. The filled circles denote the type of equilibria: unstable node in red, stable node in green and a saddle point in yellow.}
\label{Fig: dynamics focusing 4}
\end{figure}

In addition to the bend radius, particle size and aspect ratio, the initial location of the particle inside the cross-section also influences the focusing dynamics for particles. Figure~\ref{Fig: dynamics focusing 4}(a)-(c) show filled contour plots of the number of turns needed to focus a particle of size $\tilde{a}=0.05$ to an equilibrium position based on its initial cross-sectional location in a square cross-section at different bend radii. For relatively large bend radii, we observe that focusing happens quicker along a horizontal band and a vertical band connecting the two stable equilibria, while the focusing takes longer if the particles start along the diagonal lines joining the saddle points. At intermediate bend radii, where we only have two stable nodes, we find that the cross-section region is divided into three different time scales. Particles starting near the left side of the duct focus quickly in about half a turn to the stable node near the left edge whereas particles starting from the right side of the duct take around two turns to focus on the stable node near the right edge. This can be attributed to the presence of the two saddle points which slow down the focusing to the stable node along the slow manifold near the right edge. Particles starting in the middle region of the duct can take as long as three and a half turns to focus. This is because particles starting in this region first quickly migrate onto the slow manifold, and then slowly migrate along the slow manifold to the stable node near the left edge. Hence, we see a disparity in the number of turns taken to focus based on the initial location of the particle inside the duct. This is further enhanced at smaller bend radii as shown in Figure~\ref{Fig: dynamics focusing 4}(c) where, in the vicinity of the stable equilibrium point near the left edge, the particle can focus in about one turn while particles starting near the saddle point on the right may take up to sixteen turns to focus. We note that a contour plot similar to Figure~\ref{Fig: dynamics focusing 4}(c) was also obtained by Ha et al.~\cite{Kyung2021} using their simplified ZeLF model where they plotted the axial distance and the time required for a small particle to focus in a square cross-section.

For a larger particle of size $\tilde{a}=0.15$ in a square cross-section at large bend radii (see Figure~\ref{Fig: dynamics focusing 4}(d)) we observe similar focusing as for a smaller particle in Figure~\ref{Fig: dynamics focusing 4}(a), where focusing happens quicker along the lines joining the unstable nodes to stable nodes as compared to the lines joining the unstable nodes to saddle points. At intermediate bend radii where the equilibria in Figure~\ref{Fig: dynamics focusing 4}(e) are similar to those in Figure~\ref{Fig: dynamics focusing 4}(b), we find a qualitative change in the focusing contours. Here, as opposed to Figure~\ref{Fig: dynamics focusing 4}(b), we find that particles starting inside the triangular region on the right side of the duct require more turns to focus than those starting in the middle region of the cross-section. At smaller bend radii, we again find similar focusing behavior between the two particle sizes where focusing is significantly slower for initial conditions near a saddle point.

To understand the effect of the cross-section, we also plotted to contours for a larger particle of size $\tilde{a}=0.15$ in a rectangular $2\times1$ cross-section. Here at large bend radii, we see qualitatively similar focusing contours as the square cross-section for the same particle size. For an intermediate bend radius shown in Figure~\ref{Fig: dynamics focusing 4}(h), we find that focusing is quicker along a curved band joining the unstable node and the two stable nodes. Once the unstable node and the saddle point near the left edge vanish in a saddle-node bifurcation at small bend radii, the curved band of rapid focusing splits into two as shown in Figure~\ref{Fig: dynamics focusing 4}(i).

Hence, we see that the initial location of a particle can have a significant impact on the number of turns required to achieve focusing as the migration slows down in the vicinity of saddle points as well as along the slow manifold. 

\section{Conclusions}\label{sec: conclusion}

%\begin{enumerate}
%    \item In future work mention about extending the analysis to spiral ducts which we are also currently investigating
%    \item In future work, I can also include extending the present bifurcation analysis to symmetrical and asymmetrical trapezoid and other cross section which I will probably explore in future
%\end{enumerate}

We have investigated the bifurcations in particle equilibria for particles suspended in flow through curved ducts with rectangular cross-sections. We observed a rich variety of bifurcations, such as saddle-node, pitchfork and Hopf bifurcations, that take place in particle equilibria as a function of the bend radii and particle size for a square, $2\times1$ rectangular and $1\times2$ rectangular cross-sections. At large bend radii, we observed stable nodes, unstable nodes and saddle points across all particle sizes and aspect ratios. The disparity in magnitudes of the eigenvalues of these fixed points at large bend radii results in the emergence of a slow manifold. At small bend radii, we observed unstable spiral with limit cycles for particles in a $1\times2$ rectangular cross-section and a small particle in a square cross-section, while for all the other particle sizes and aspect ratios considered in this work, we observed stable spirals. Investigating the variations in the stable particle equilibria location as a function of the bend radii for different particle sizes and aspect ratios showed a good separation between the smaller particle sizes. The best separation was achieved for wider rectangular ducts across a broad range of bend radii that are typical in experiments.

We also explored the effects of bend radii, particle size, aspect ratio and the initial particle location within the cross-section, on the particle focusing dynamics. We observed that although particles focus more quickly to their equilibria at smaller bend radii compared to larger bend radii, one requires more turns for particles to completely focus at smaller bend radii. Comparisons of two different particle sizes initiated at similar locations with similar equilibria revealed that larger particles can focus using a slightly smaller angular distance of a curved duct compared to the smaller particle. Simulating the particle dynamics with different aspect ratios showed that low aspect ratio rectangular ducts can focus using less turns compared to high aspect ratio ducts. Lastly, we observed the number of turns required for particles to focus to their equilibria can be significantly influenced by their starting location inside the cross-section. All these preliminary observations of the dynamics of particle focusing can be extended and studied in more detail which will be useful in comprehensive designing of inertial microfluidic devices.

The work presented in this paper can be extended in several directions. One extension of the present could be to investigate the dynamics of inertial focusing in spiral ducts where the bend radius, and hence the location of the particle equilibria, changes with angle. A slowly changing bend radius would result in particles following the corresponding fixed points as they move, but it would be interesting to investigate the non-equilibrium dynamics when the bend radius crosses a bifurcation or changes swiftly compared to the time it takes for the particle to focus. We plan to investigate this further in a future work. Moreover, it would be interesting to extend the present work to non-rectangular cross-sections. For example, trapezoidal cross-sections are known to be more effective at particle focusing and hence it would be valuable to systematically understand the equilibrium focusing points and the various bifurcations that take place in such geometries.

Different system parameters such as channel and cross-section geometries, flow rates and particle size are likely to result in different particle focusing locations, but our current understanding of inertial focusing does not provide us with an intuition to predict particle focusing in arbitrary settings. Currently, the innovations in inertial microfluidics are primarily explored more efficiently using experiments. We hope that the numerical results presented in this paper motivate further development of analytical and numerical tools that can be efficiently used to decouple various physical effects and facilitate a rapid exploration of a large parameter space. 

%\appendix
%\section{An example appendix} 

\section*{Acknowledgments}
This research is supported under Australian Research Council’s Discovery Projects funding scheme (project number DP160102021 and DP200100834). The results were computed using supercomputing resources provided by the Phoenix HPC service at the University of Adelaide and the Raapoi HPC service at Victoria University of Wellington.

\bibliography{references}
\bibliographystyle{siamplain}

\end{document}